# Transmission of Light in Crystals with different homogeneity: Using Shannon Index in Photonic Media

Michele Bellingeri[1], Stefano Longhi[2], Francesco Scotognella[*,2],

[1] Dipartimento di Scienze Ambientali, Università di Parma, Parco Area delle Scienze, 33/A 43100 Parma, Italy
[2] Dipartimento di Fisica, Politecnico di Milano, piazza Leonardo da Vinci 32, 20133 Milano, Italy
* Corresponding author: francesco.scotognella@gmail.com, Phone: +39 02 2399 6091, Fax: +39 02 2399 6126

Light transmission in inhomogeneous photonic media is strongly influenced by the distribution of the diffractive elements in the medium. Here it is shown theoretically that, in a pillar photonic crystal structure, light transmission and homogeneity of the pillar distribution are correlated by a simple linear law once the grade of homogeneity of the photonic structure is measured by the Shannon index, widely employed in statistics, ecology and information entropy. The statistical analysis shows that the transmission of light in such media depends linearly from their homogeneity: the more is homogeneous the structure, the more is the light transmitted. With the found linear relationship it is possible to predict the transmission of light in random photonic structures. The result can be useful for the study of electron transport in solids, since the similarity with light in photonic media, but also for the engineering of scattering layers for the entrapping of light to be coupled with photovoltaic devices.

## 1. Introduction

Propagation of electromagnetic waves in complex and structured photonic media is a topic of major importance for the comprehension of some general properties of transport phenomena and for a better understanding of the transport of electronic in solids owing to well-known analogies between electronic and photonic transport. [1-3] Complex dielectric structures show variations of the refractive index on a length scale comparable to the wavelength of light. In ordered structures, namely in photonic crystals, for a certain range of energies and certain wave vectors, light is not allowed to propagate through the medium [4-6]. This behaviour is very similar to the one of electrons in a semiconductor, where energy gaps arise owing to the periodic crystal potential at the atomic scale. Photonic crystals are present in nature or can be fabricated through a wide range of techniques, with the dielectric periodicity in one, two and three dimensions [7-9]. Nowadays, these materials are extensively studied since they find application in several fields, including photonics for low threshold laser action, high bending angle waveguide, super-prism effect, sensors and optical switches. [10-15] The optical properties of photonic crystals, as for example the transmission of light, can be predicted by several efficient mathematical methods [6,16-19], but for less homogeneous structures these calculations become very cumbersome. For a better comprehension of the optical properties of such complicated systems, simple methods that are not time consuming can be very useful. Recently, concepts and methods widely used in statistics have been successfully applied to explain light transport phenomena in Lévy glasses [20].

In this work, we have analyzed the light transmission properties of two-dimensional photonic media, studied by the use of a finite element method, and we have demonstrated a simple scaling law between transmission of light over a wide range of wavelengths and the distribution of the diffractive elements in the photonic lattice, the grade of homogeneity of the structure being quantified by the Shannon index, commonly employed in statistics and information theory [20]. In particular, we have shown that the transmission of light increases linearly by increasing the Shannon index, i.e. by increasing the homogeneity of the distribution of pillars in the crystals.

## 2. Outline of the Method

We consider a well-known structure of two-dimensional photonic crystal: a square lattice of dielectric circular pillars [6]. The diameter of the column $d$ is set to 75 nm and the lattice constant $a$ is 300 nm (Fig. 1), the pillars are made of Titanium dioxide ($n_T$ = 2.45) and the matrix where the pillars are embedded is Silicon dioxide ($n_S$ = 1.46). Note that, for such a geometrical setting $n_T d \sim$

$n_S$ (*a-d*) is satisfied [6]. Starting from this regular structure, keeping constant the number of pillars throughout the crystals, a number of less regular crystals have been synthesized by concentrating more and more pillars in certain unit cells of the original crystal. In this way, some unit cells of the crystal have no pillars. Figure 2 shows a few representative realizations of the different crystals configurations. For clarity, the different crystals are named corresponding to the number of pillars per cell (1, 2, 3, …, 16).

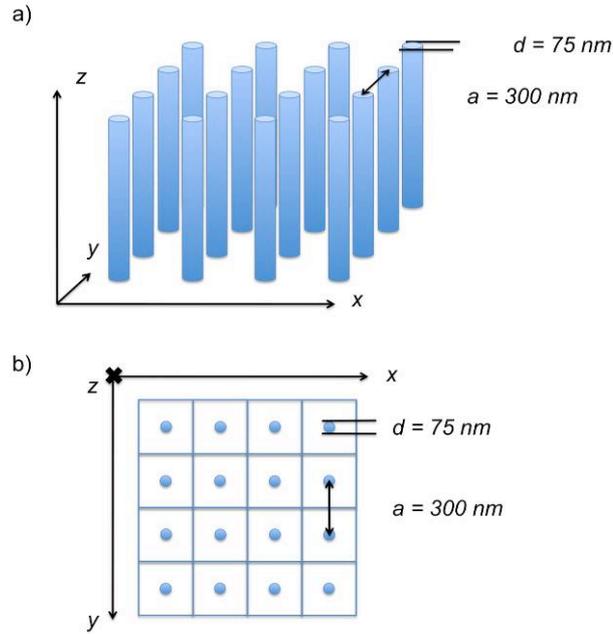

**Figure 1.** a) Schematic of a square lattice of dielectric pillars; b) The diameter of the column *d* is 75 nm and the lattice constant *a* is 300 nm.

The pillar distributions in the different photonic structures are set up in order to have a selected homogeneity, that we describe herein with the so called Shannon index. In several fields of science, the diversity is correlated to a Shannon index [21,22]. The Shannon-Wiener *H'* index is a diversity index used in statistics, defined as

$$H' = -\sum_{j=1}^{s} p_j \log p_j \qquad (1)$$

where $p_j$ is the proportion of the j-fold species and *s* is the number of the species. This index is widely used in statistics and ecology as an evenness measurement and in physics in the field of information theory [21-24]. Dividing *H'* by *log(s)* we have normalized the index constraining it within the range (0,1). We used the normalized Shannon index (i.e. $0 \leq H' \leq 1$) as a measurement of the homogeneity of the transmission medium: in this study $p_j$ indicates the proportion of pillars belonging to the *j*-fold cell and *s* the total number of cells. In an analytical way, *H'* is the highest when the medium has perfect uniform structure and is the lowest when the medium is completely scalar (in our case the most aggregated it is when all the pillars are in one cell, a limit structure that we have not used in our experiment since it is not possible, in our two dimensional model, to put more than sixteen pillars in a cell).

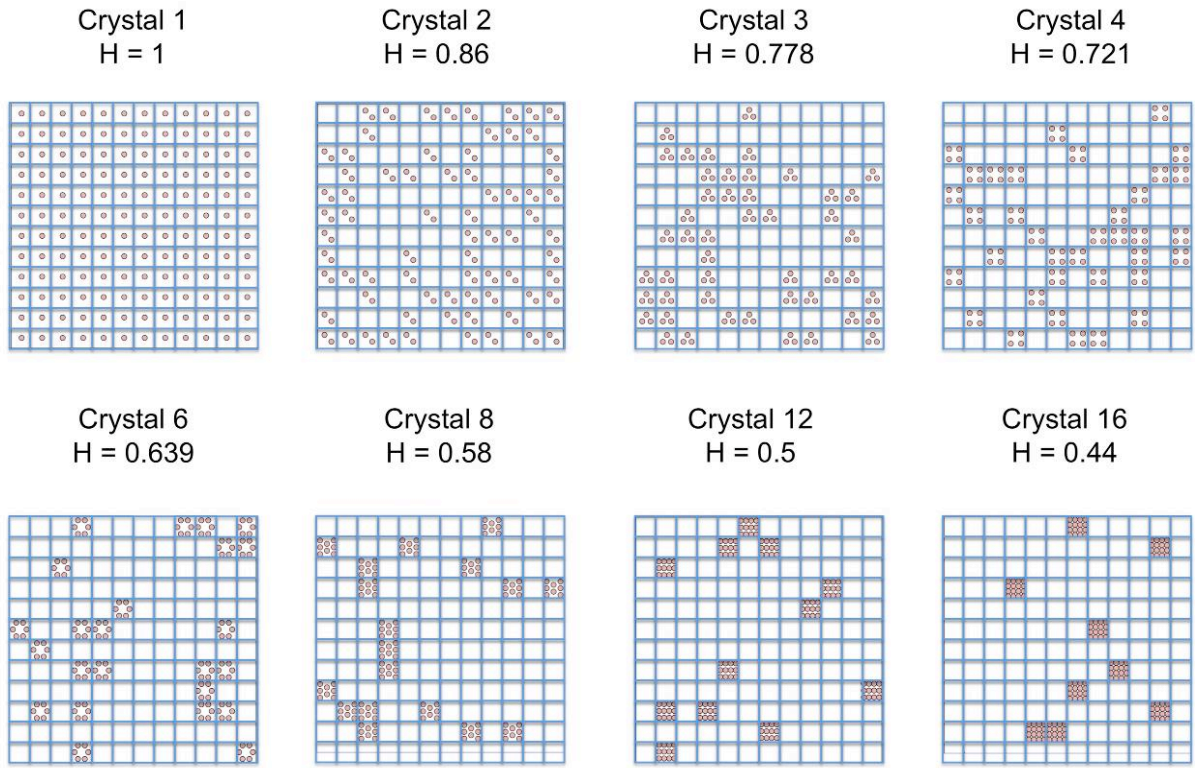

**Figure 2.** Examples of realizations of photonic structures with different pillar distributions. Each scheme is correlated to a determined Shannon index.

To set up the crystals, we consider a 12x12 photonic crystal cells and we start getting one pillar inside each cell to build the perfect uniform structure: a regular lattice of Titanium dioxide, i.e. crystal 1 in Figure 2. This first structure is the square lattice of dielectric pillars [6], the crystal with the maximum evenness [22] in which *H'* is equal to 1. At second step, we build a less uniform crystal by aggregating two pillars in each cell, thus a certain number of cells becomes empty. Crystal 2 in Fig. 2 gives a possible realization as an illustrative example. At third step we concentrate three pillars in each cell leaving an increasing number of cells without pillars. We continue the concentration of diffractive elements with this method for the following crystals until to reach the most aggregated structure we can make, by taking into account the maximum number of pillars which can be contained in a cell respecting the optical distance between pillars (in our case 16). Thus, the aggregation method makes crystals in which all no empty cells have the same number of pillars, i.e. in Crystal 1 every no empty cell has one pillar, in Crystal 2 every cell has 2 pillars, in Crystal 3 every no empty cell has 3 pillars, *etc...* For each crystal *n*, ten different realizations have been considered by allocating in a random fashion the (12×12)/*n* clusters among the available 12×12 cells of the original crystal. In this way, the ten realizations of each crystal *n* have the same Shannon index. As an example, the different distributions for Crystal 2 are depicted in Figure S1 in the Supporting Information.

Since for a crystal where the pillars aggregate in clusters in some unit cells (and the other unit cells become consequently empty) the Shannon index decreases, we have obtained a set of crystals, from the most uniform to the most aggregated, in inverse proportion to *H'*.

For the calculation of the light transmission of the photonic structures through the finite element method, we assumed a TM-polarized field and used the scalar equation for the transverse electric field component $E_Z$

$$\left(\partial_x^2 + \partial_y^2\right)E_Z + n^2 k_0^2 E_Z = 0 \qquad (2)$$

where *n* is the refractive index distribution and $k_0$ is the free space wave number [6,25]. As input field, a plane wave with wave vector *k* directed along the *x* axis has been assumed. Scattering boundary conditions in the *y* direction has been used.

## 3. Results and Discussion

We have calculated, by the finite element method [6,25], the transmission of light in a broad range of wavelength (450 – 1400 nm) for different crystals, starting from a perfect uniform one, in which the evenness and the Shannon index is at the maximum, to the most aggregated structures.

For each crystal the Shannon index, calculated according to Eq. 1, is reported in Table 1 as well as above the crystal schemes depicted in Figure 2. The corresponding transmission spectra have been calculated in the 450 – 1400 nm wavelength range, i.e, including visible and part of the near infrared (NIR) radiation. This range of wavelengths can be assimilated to the emission spectrum of a supercontinuum source.

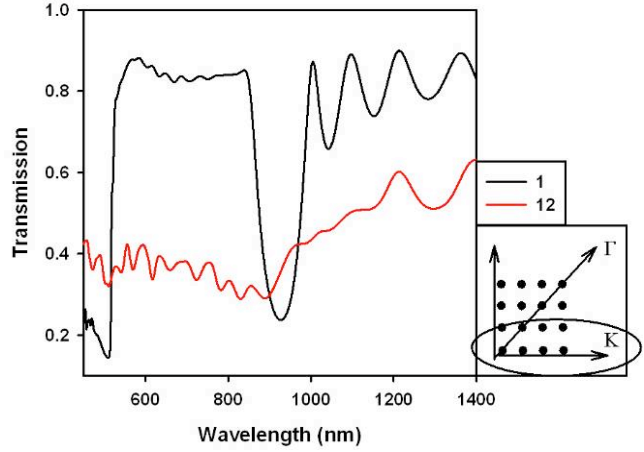

**Figure 3.** Calculated transmission spectra for Crystal 1 and Crystal 12, along the crystallographic direction *K*.

In Figure 3 the calculated transmission spectra, along the crystallographic direction *K*, of crystals 1 and 12 (i.e., for one of the ten distributions that have been designed for Crystal 12) are depicted. Crystal 1 is a regular photonic crystal that shows a photonic band gap centred at 925 nm according to the Bragg Snell law $\lambda_{Bragg} = n_{eff} \Lambda$, where $\lambda_{Bragg}$ is the centre wavelength of the stop band, $n_{eff}$ is the effective refractive index of the lattice and $\Lambda$ the spatial period (in our case, $\Lambda = a = 300$ nm). The transmission spectrum of Crystal 12 does not display a photonic band gap. On the other hand, different peaks over all the wavelengths are present and the amount of light transmitted through the crystal is relatively low in all the range 450 – 1400 nm.

| Crystal | Shannon index | Average Transmission 450-1400 nm | Error (Standard Deviation) |
|---------|---------------|----------------------------------|----------------------------|
| 1 | 1 | 1 | - |
| 2 | 0.86 | 0.8674 | 0.0224 |
| 3 | 0.778 | 0.8086 | 0.0243 |
| 4 | 0.721 | 0.8429 | 0.0297 |
| 6 | 0.639 | 0.7908 | 0.0426 |
| 8 | 0.58 | 0.8054 | 0.0604 |
| 12 | 0.5 | 0.7043 | 0.0727 |
| 16 | 0.44 | 0.7436 | 0.0772 |

**Table 1.** Shannon index and transmission of light in the 450 – 1400 nm spectral range for each crystal. The last column gives the standard deviation of the average transmission for the ten different realizations.

For all the crystals, the fractional power transmission, averaged over the entire spectral range, has been computed and correlated to the Shannon indices (see Table 1). In Figure 4 we plot these values as a function of the Shannon index of the corresponding crystal. The behaviour that we have found is a clear increase of light transmission in the range 450 – 1400 nm by increasing the Shannon index: this means that the more the pillars are clustered in certain unit cells the less the light is transmitted through the crystal. It is noteworthy that this trend is in good agreement with a linear behaviour ($R^2$ = 0.845, coefficient of determination). This result could be counterintuitive since one should expect that, by enlarging portions of space without diffractive elements in the crystal, the light could be more easily transmitted. It should be also noticed that the linear increase of average transmission with Shannon index turns out to be rather insensitive to slight variations of the spectral range. For example, if we integrate the intensity of light transmission in the 525 – 1325 nm spectral range (instead of 450 – 1400), we observe the same linear trend of the light transmission as a function of the photonic structure homogeneity.

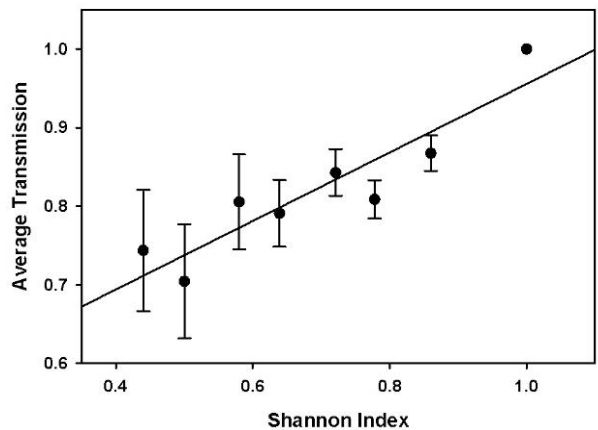

**Figure 4.** Average transmission of the crystals as a function of the Shannon index.

One could ask whether the linear trend shown in Fig. 4, and in particular the values of average transmission predicted by the linear fit of Fig. 4, could be extended to the case where the pillar cluster sizes are not uniform. To this aim, we have performed a further *in silico* experiment, where we have selected a crystal with a random but skewed distribution of pillars in the cells. To design this crystal, we have assigned pillars in cells by a fitness model [26]. We have used this model in order to make a crystal space with skewed clusters size without benchmark distribution. Thus, we have obtained a random crystal in which the clusters size distribution (i.e. pillars for cells distribution) is skewed. This crystal has no empty cells with various number of pillars and therefore it differs with respect to the set of crystals used in the first experiment.

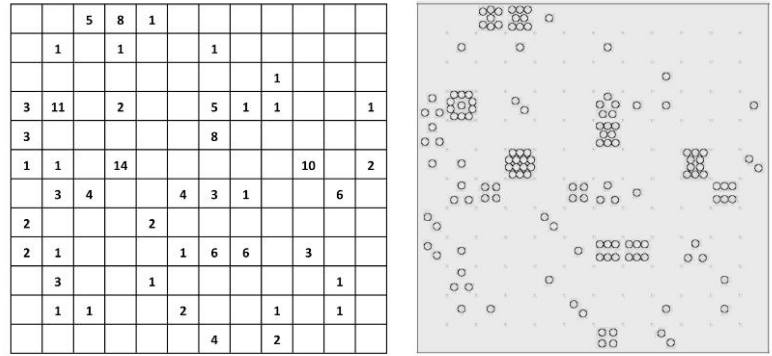

**Figure 5.** Scheme of Crystal R.

Figure 5 shows the scheme of the crystal with the random skewed distribution of pillars (here called crystal R). Crystal R has a Shannon index of 0,6931. Starting from this crystal, as we have done for each crystal in the former set of crystals, we have designed ten different replicates by varying the position of the pillar clusters (i.e. ten cell permutations), but by keeping constant the cluster size set and, consequently, the Shannon index of the crystals. In Figure S2 in the Supporting Information the ten differents crystals are shown. The resulted average transmission of crystal R (performed with the same method used for crystals 1,2,3 …) gives a value of 0.8059 ± 0.0334, while the value extracted from the linear fit of the previous experiment, for a Shannon index of 0.6931, is 0.8218 ± 0.0380. This indicates that, for the example investigated, the average transmission of a photonic structure with random pillar cluster sizes can be predicted by the linear relationship between average light transmission and Shannon index that we have found. Such a result might suggest that the Shannon index could provide a global parameter to approximatively predict light transmission even for randomized cluster size photonic structures.

## 3. Conclusions

In conclusion, in this *in silico* experiment we have found that the evenness of the photonic crystal can affect its transmission properties in a broad range of wavelengths (about ±500 nm around the centre of the photonic band gap of crystal 1), with a linear trend. Our study indicates that a coarse-grained behaviour of light transmission, in the case of two dimensional crystals with a disorganized cluster distribution of pillars, can be estimated from a simple computation of the Shannon index as the parameter that measures the grade of homogeneity of the structure. Owing to the analogy between photonic transport in periodic media and electronic transport in semiconductors, our results could be of interest beyond optics, providing useful insights into the transport properties of electrons in inhomogeneous semiconductor lattices and superlattices. From an applied point of view, our results can be useful for the engineering of scattering layers for light coupling and trapping enhancement in photovoltaic devices [27] and for the study of light scattering in diffusive media for diagnostic imaging [28].

**Acknowledgements**

The authors would like to acknowledge Prof. Guglielmo Lanzani, Angelo Monguzzi and Riccardo Gatti for help and discussions.

**Supporting Information**

**Figure S1.** Different distributions for Crystal 2

```
      [,1] [,2] [,3] [,4] [,5] [,6] [,7] [,8] [,9] [,10] [,11] [,12]
 [1,]   0    0    2    2    0    2    2    2    0    2     2     2
 [2,]   0    0    2    0    0    0    0    0    2    2     2     0
 [3,]   2    2    0    0    0    2    0    2    0    0     0     2
 [4,]   0    2    0    2    2    0    2    2    0    0     0     2
 [5,]   2    2    0    0    0    0    0    2    2    2     2     2
 [6,]   2    2    0    0    0    2    0    2    0    0     2     0
 [7,]   2    0    0    0    0    0    2    2    2    0     2     0
 [8,]   2    0    0    0    2    0    0    2    0    0     0     2
 [9,]   2    2    2    0    2    0    0    2    2    2     0     2
[10,]   0    0    2    0    0    2    2    2    2    0     2     2
[11,]   2    0    0    0    2    0    2    2    0    2     0     0
[12,]   2    2    2    2    2    0    0    2    2    0     2     2
```

```
      [,1] [,2] [,3] [,4] [,5] [,6] [,7] [,8] [,9] [,10] [,11] [,12]
 [1,]   2    2    0    2    0    0    2    2    2    0     0     2
 [2,]   2    2    0    0    0    2    0    0    2    2     2     2
 [3,]   2    0    0    2    0    0    2    2    2    0     0     0
 [4,]   0    0    2    2    2    2    0    0    0    2     2     2
 [5,]   2    0    2    0    2    2    2    0    2    0     0     0
 [6,]   2    0    2    2    2    2    0    2    2    2     0     0
 [7,]   0    2    0    2    0    0    0    2    0    2     0     0
 [8,]   2    0    2    2    2    0    2    2    0    2     0     0
 [9,]   2    2    0    2    2    0    2    0    2    2     2     0
[10,]   0    0    0    0    0    0    2    0    2    2     2     0
[11,]   2    0    0    2    2    0    0    2    0    0     0     0
[12,]   0    2    0    0    2    2    0    0    2    0     0     2
```

```
      [,1] [,2] [,3] [,4] [,5] [,6] [,7] [,8] [,9] [,10] [,11] [,12]
 [1,]   0    0    0    0    0    0    2    0    2    0     0     0
 [2,]   2    2    2    2    0    2    2    0    2    2     0     2
 [3,]   2    2    0    2    0    2    2    0    2    2     2     2
 [4,]   2    0    2    0    2    2    2    0    0    2     2     2
 [5,]   2    0    0    0    2    2    0    2    0    2     0     0
 [6,]   2    0    2    0    2    2    0    2    2    0     0     2
 [7,]   0    0    0    2    2    0    0    2    0    0     0     0
 [8,]   0    2    2    0    2    0    0    0    2    2     2     0
 [9,]   0    2    2    0    2    2    0    2    2    0     2     2
[10,]   2    0    0    2    0    0    0    0    2    2     2     0
[11,]   0    2    2    2    2    0    0    2    0    0     0     0
[12,]   0    0    0    2    2    2    0    2    0    0     0     2
```

```
      [,1] [,2] [,3] [,4] [,5] [,6] [,7] [,8] [,9] [,10] [,11] [,12]
 [1,]   2    2    2    2    0    0    0    0    2    2     2     2
 [2,]   0    0    2    0    2    0    2    0    2    0     2     0
 [3,]   2    0    2    0    2    2    2    0    2    0     0     0
 [4,]   0    0    0    0    2    0    2    2    0    2     0     2
 [5,]   0    0    0    0    0    0    0    0    0    2     0     0
 [6,]   0    2    0    0    2    2    2    0    0    2     2     0
 [7,]   0    0    2    0    2    2    0    2    0    2     2     0
 [8,]   2    2    2    2    2    2    2    0    2    0     0     0
 [9,]   2    0    2    2    2    2    0    2    2    0     0     0
[10,]   2    0    2    2    0    2    0    2    2    2     2     2
[11,]   2    0    0    0    0    0    0    2    2    2     0     2
[12,]   0    0    2    2    2    2    0    2    0    0     2     2
```

```
      [,1] [,2] [,3] [,4] [,5] [,6] [,7] [,8] [,9] [,10] [,11] [,12]
 [1,]   0    0    0    0    2    0    2    2    2    0     2     0
 [2,]   2    0    0    0    0    0    0    0    2    0     2     2
 [3,]   0    2    2    0    0    0    0    2    0    0     2     0
 [4,]   2    2    2    2    0    2    2    0    2    2     2     0
 [5,]   0    0    0    0    0    2    2    2    2    2     0     0
 [6,]   2    0    2    2    0    2    0    0    0    2     2     0
```

```
 [7,]   2  2  0  0  0  0  2  0  0  0  2  0
 [8,]   2  2  2  2  2  2  0  2  0  2  0  0
 [9,]   2  2  2  2  2  0  2  2  2  0  0  0
[10,]   2  2  0  2  2  2  2  2  0  2  0  0
[11,]   2  0  0  2  0  2  2  2  0  0  0  2
[12,]   0  2  2  2  2  0  2  0  0  0  0  0

      [,1] [,2] [,3] [,4] [,5] [,6] [,7] [,8] [,9] [,10] [,11] [,12]
 [1,]   2   0   0   2   2   0   2   0   0   0    2    0
 [2,]   0   0   2   0   2   2   2   2   2   0    2    0
 [3,]   2   0   0   2   2   2   2   0   0   2    0    0
 [4,]   0   2   0   2   2   2   0   2   0   2    0    2
 [5,]   2   2   2   2   2   0   0   0   0   0    0    0
 [6,]   2   2   0   0   0   0   2   2   0   0    0    2
 [7,]   2   0   2   0   2   0   2   2   0   2    0    2
 [8,]   2   2   0   2   0   0   2   0   0   2    2    2
 [9,]   2   2   0   0   0   0   2   0   0   2    0    0
[10,]   2   0   2   2   0   2   2   2   2   0    0    0
[11,]   2   2   2   0   2   0   0   2   0   0    0    2
[12,]   0   2   0   2   0   2   2   2   0   2    0    0

      [,1] [,2] [,3] [,4] [,5] [,6] [,7] [,8] [,9] [,10] [,11] [,12]
 [1,]   0   2   2   0   2   0   2   0   0   2    2    0
 [2,]   0   2   0   0   2   2   0   0   2   0    2    2
 [3,]   0   2   0   2   2   0   2   2   2   2    0    0
 [4,]   0   0   2   2   0   0   2   2   2   2    2    0
 [5,]   2   0   0   2   0   2   2   0   0   0    0    0
 [6,]   2   0   0   2   2   2   2   0   2   2    2    2
 [7,]   0   0   0   0   0   0   0   2   2   0    0    0
 [8,]   0   2   2   0   0   2   2   0   0   2    0    2
 [9,]   2   2   0   2   2   2   2   2   0   2    0    0
[10,]   2   0   0   0   2   2   0   0   2   2    0    0
[11,]   2   0   2   0   2   2   2   0   0   0    0    0
[12,]   2   0   0   2   0   0   2   0   2   2    2    2

      [,1] [,2] [,3] [,4] [,5] [,6] [,7] [,8] [,9] [,10] [,11] [,12]
 [1,]   0   0   2   0   0   2   2   2   0   0    2    0
 [2,]   2   2   0   0   0   0   2   0   0   2    0    0
 [3,]   2   2   2   2   0   0   2   0   2   2    2    0
 [4,]   0   0   2   0   0   0   0   0   0   0    0    0
 [5,]   0   2   0   2   0   0   2   2   2   0    2    2
 [6,]   2   2   2   0   2   2   2   2   2   2    2    0
 [7,]   2   2   2   0   0   0   2   2   0   2    0    0
 [8,]   0   2   0   2   0   2   2   2   2   0    2    0
 [9,]   0   2   0   0   0   0   2   0   2   0    2    2
[10,]   0   0   0   2   2   0   2   0   2   2    2    2
[11,]   0   2   0   0   0   0   0   2   2   2    2    2
[12,]   2   0   0   2   2   0   2   2   2   0    0    0

      [,1] [,2] [,3] [,4] [,5] [,6] [,7] [,8] [,9] [,10] [,11] [,12]
 [1,]   0   2   0   2   0   2   0   2   2   0    0    0
 [2,]   0   2   2   2   0   2   0   2   0   0    0    2
 [3,]   0   2   0   0   2   2   0   0   0   0    0    0
 [4,]   0   2   2   2   2   2   0   0   2   2    0    0
 [5,]   2   2   0   0   2   2   0   2   2   0    0    2
 [6,]   2   2   0   2   0   2   0   2   0   0    0    0
 [7,]   2   0   0   0   2   0   2   0   0   0    0    2
 [8,]   0   0   2   0   0   0   2   0   0   0    0    2
 [9,]   0   2   0   0   0   2   2   2   0   2    2    2
[10,]   2   2   2   0   2   0   2   2   2   2    2    0
[11,]   0   2   2   0   0   2   2   2   2   2    2    2
[12,]   0   2   0   0   2   2   2   2   2   0    2    2

      [,1] [,2] [,3] [,4] [,5] [,6] [,7] [,8] [,9] [,10] [,11] [,12]
 [1,]   2   2   0   0   0   2   2   0   2   0    2    0
 [2,]   2   0   2   2   2   0   0   0   2   0    2    2
 [3,]   0   0   2   2   0   0   2   0   0   0    2    2
 [4,]   2   2   0   0   2   2   0   2   2   0    0    2
```

```
 [5,]  2  2  2  0  0  2  0  0  2  2  0  0
 [6,]  0  2  2  2  0  2  0  0  0  2  0  0
 [7,]  2  0  2  0  2  0  0  0  0  2  0  2
 [8,]  2  2  0  2  2  0  2  0  0  0  0  2
 [9,]  0  2  0  2  2  0  0  2  2  2  0  0
[10,]  0  0  0  0  2  2  2  0  2  2  2  2
[11,]  0  2  0  2  2  2  0  0  0  0  0  2
[12,]  2  0  0  0  2  0  2  2  0  2  2  2
```

**Figure S2.** Different distributions for Crystal R.

```
       [,1] [,2] [,3] [,4] [,5] [,6] [,7] [,8] [,9] [,10] [,11] [,12]
 [1,]   0    0    3    3    5    2    1    0    0    1    1    0
 [2,]   2    1    0    0    0    0    0    0    3    0    0    0
 [3,]   0   14    1    2    0    4    0    0    0    0    0    1
 [4,]   0    0    0    0    1    0    1    3    0    0    8    1
 [5,]   0    0    1    0    0    0    6    0    1    0    0    0
 [6,]   0    0    0    0    1    0    0    1    1    0    0    3
 [7,]  10    0    0    0    2    0    0    2    0    8    0    0
 [8,]   4    0    4    0    1    1    3    0    0    0    0    0
 [9,]   2    0    0    0    0    1    0    0    0   11    0    0
[10,]   0    0    0    0    0    0    0    0    0    0    0    0
[11,]   0    0    0    1    0    0    7    0    0    1    0    6
[12,]   6    0    0    0    0    0    0    0    0    0    2    0
```

```
       [,1] [,2] [,3] [,4] [,5] [,6] [,7] [,8] [,9] [,10] [,11] [,12]
 [1,]   0    0    0    0    4    0    1    0    2    1    0    0
 [2,]   6    0    1    4    0    0    0    0    0    0    3    0
 [3,]   0    0    0    1    0    0    0    0    0    0    2    1
 [4,]   1    0    0    0    2    0    1    0    1    1    0    1
 [5,]   2    0    2    0    3    0    0    0    0    0    0    0
 [6,]   0    7    0    3    0    1    0    0    2    6    3    8
 [7,]   0    0    0    0    0    1    0    0    3    0    0    0
 [8,]   0    6    1    0    2    0    0    0    0    0    0    0
 [9,]   0    0    0   14    0    0    0    0    0    0    0    1
[10,]   0    0   10    0    1    0    0    0    0    1    0    0
[11,]   0    0    0    0    0    0    0    1    0    0    0    0
[12,]   0    1    0    3    0    0    4    5    1    8   11    0
```

```
       [,1] [,2] [,3] [,4] [,5] [,6] [,7] [,8] [,9] [,10] [,11] [,12]
 [1,]   0    0    1    1    0    0    0    0    0    0    0    0
 [2,]   0    1    6    0    0    1    1    1    1    0    0    2
 [3,]   0   14    0    1    8    0    0    2    0    0    0    0
 [4,]   0    3    0    0    0    0    3    0    0    0    0    0
 [5,]   0    1    0    0    3    0    0    0    3    0    1    0
 [6,]   7    1    0    0    0    0    2    0    1    3    0    1
 [7,]   0    0    0    0    0    0    0    0    0    0    5    0
 [8,]   0    0    0    0    0    2    2   10    0    0    0    1
 [9,]   0    0    0    0    1   11    8    0    1    0    0    0
[10,]   0    0    6    4    0    0    0    0    0    0    2    0
[11,]   2    6    0    0    0    4    0    4    0    0    1    0
[12,]   0    0    0    0    0    3    1    0    1    0    0    0
```

```
       [,1] [,2] [,3] [,4] [,5] [,6] [,7] [,8] [,9] [,10] [,11] [,12]
 [1,]   0    6    0    0    0    0    0    2    0    0    0    0
 [2,]   0    1    0    0    0    0    1    4    0    0    1    3
 [3,]   0    0    0   10    0    0    2    0    0   14    0    0
 [4,]   0    0    0    1    0    0    0    0    1    3    1    0
 [5,]   0    3    0    5    3    0    1    0    0    0    1    0
 [6,]   0    1    0    0    0    0    0    0    0    1    0    4
 [7,]   0    0    2    0    0    0    0    1    1    0    0    0
 [8,]   0    0    0    0    2    0    0    0    0    0    0    8
 [9,]   0    8    0    2    0    0    6    4    0    1    1    6
[10,]   0    1    0    0    0    0    0    2    7    3    1    0
[11,]   0    0    0    0    2    0    1    1    0    0    1    0
[12,]   0    0    3    0    0    0    0    0    0    0   11    0
```

```
      [,1] [,2] [,3] [,4] [,5] [,6] [,7] [,8] [,9] [,10] [,11] [,12]
 [1,]    0    1    0    0    0    0    0    1    0     0     3     0
 [2,]    0    0    2    0    0    1    0    1    0     0     1     1
 [3,]    0    0    0    4    0    2    0    0    0     0     0     0
 [4,]    8    1    0    0    0    0    0    0    0     0     0     7
 [5,]    1    0    0    1    0    0    0    0    0     0     1     0
 [6,]    0    0    0    0    0    1    0    0    0     0     0     0
 [7,]    2    0    0    5    1    1    0    0    1     3    11     0
 [8,]    0    0    0    8    0    0    0    0    0     3     0     0
 [9,]   14    0    0    0    0    0   10    0    2     1     1     0
[10,]    0    0    4    3    1    0    0    6    0     0     3     4
[11,]    0    2    0    0    0    0    0    0    0     2     0     0
[12,]    0    0    1    6    6    0    3    0    0     2     1     0

      [,1] [,2] [,3] [,4] [,5] [,6] [,7] [,8] [,9] [,10] [,11] [,12]
 [1,]    1    0    0    0    0    6    1    0    0     0     0     8
 [2,]    0    5    8    1    0    0    0    3    0     1     0     2
 [3,]    4    2    0    0    7    0    0    0    0     0     3     3
 [4,]    0    0    0    0    1    0    0    0    1     0     0     0
 [5,]    0    0    0    0    0    0    0    0    1     0     0     0
 [6,]    0   14    0    0    0    0    1    0    1     1     0     0
 [7,]    3    6    0    0    0    0    0    0    0     0     0     0
 [8,]    1    2    0    0    0    0    0    3    0     0    10     0
 [9,]    1    0    0    2    0    4    0    0    0     0    11     0
[10,]    0    0    0    1    0    0    1    4    0     0     0     1
[11,]    1    0    0    0    1    0    0    3    0     0     0     0
[12,]    0    1    0    2    6    1    0    2    0     2     0     0

      [,1] [,2] [,3] [,4] [,5] [,6] [,7] [,8] [,9] [,10] [,11] [,12]
 [1,]    0    0    0    0    1    0    0    0   10     0     0     0
 [2,]    4    6    0    1    3    2    0    0    1     5     1     0
 [3,]    0    1    0    0    2    0    1    1    0     0     1     2
 [4,]    0    0    0    0    4    0    0    0    0     0     0     0
 [5,]    0    2    0    0    0    0    0    1    0     0     0     4
 [6,]    2    0    0    1    3    0    0    0    6     1     8     0
 [7,]    0    0    1    1    0    1    0    0    2     0     0     3
 [8,]    0    1    0    0    0    6    0    0    1     0     0     0
 [9,]    0   14    3    0    3    1    0    0    0     0     0    11
[10,]    0    0    0    0    0    0    0    0    0     0     0     7
[11,]    1    1    0    0    0    0    0    0    0     0     0     0
[12,]    0    0    0    0    0    0    3    0    0     0     2     8

      [,1] [,2] [,3] [,4] [,5] [,6] [,7] [,8] [,9] [,10] [,11] [,12]
 [1,]    0    4    1    0    0    0    0    0    0     0     0     0
 [2,]    0    0    0    0    0    0    1    0    0     0     0     0
 [3,]    8    0    0    0    0    2    0    0    0     0     0     0
 [4,]    0    1    4    0    6    0    0    2    0     3     0     2
 [5,]    0    0    0    0    1    0    0    0    0     0     0     1
 [6,]    0    1    2    1    1   10    0    0    1     1     0     6
 [7,]    0    0    0    3    0    0    0   11    0     7     0     3
 [8,]    0    0    1    0    6    0    1    0    0     0     0     0
 [9,]    0    0    4    1    0    3    0    1    8     0     0     0
[10,]    0    0    0    0    0    2    0    0    1     1     0     0
[11,]    0    2    5    0   14    0    0    0    0     1     2     3
[12,]    1    0    0    0    0    0    1    0    0     3     0     0

      [,1] [,2] [,3] [,4] [,5] [,6] [,7] [,8] [,9] [,10] [,11] [,12]
 [1,]    0    0    0    0    1    0    0    0    0     0     0     0
 [2,]    1    0    0    0    1    0    4   10    0     3     0     0
 [3,]    0    0    1    4    3    0    0    1    3     0     0     2
 [4,]    2    0    6    0    0    1    0    0    0     0     0     0
 [5,]   14    0    0    0    6    0    4    0    0     0     1     6
 [6,]    0    0    1    2    0    0    2    0    3     2     0     3
 [7,]    0    0    5    1    0    1    0    0    0     0     0     0
 [8,]    0    1    0    0    0    0    1    0    0     0     0     8
 [9,]    0    1    0    0    0    0    0    0    0     0     0     0
[10,]    0   11    0    0    0    8    2    0    1     0     1     0
```

```
 [11,]   1   7   0   0   3   1   0   0   0   0   0   0
 [12,]   0   0   0   0   1   2   0   0   1   0   0   0
       [,1] [,2] [,3] [,4] [,5] [,6] [,7] [,8] [,9] [,10] [,11] [,12]
 [1,]   1   0  10   0   0   0   0   0   0   0   0   0
 [2,]   0   0   0   1   0   0   0  14   0   0   0   1
 [3,]   1   0   0   0   0   0   0   0   0   0   2   0
 [4,]   0   0   6   1   0   2   8   0   0   0   0   0
 [5,]   2   2   0   0   6   0   2   1   1   0   0   0
 [6,]   0   0   0   7   0   0   6   0   0   3   0   1
 [7,]   0   0   1  11   0   0   0   0   0   1   4   0
 [8,]   0   0   0   0   0   0   0   5   0   0   8   1
 [9,]   0   1   0   3   0   0   0   0   0   0   3   3
[10,]   0   0   4   0   0   0   3   0   3   1   0   1
[11,]   1   1   0   1   4   0   0   0   0   2   0   0
[12,]   0   0   0   0   0   0   0   2   1   1   0   0
```